\documentstyle[prl,twocolumn,epsfig,aps]{revtex}
 
\newcommand{\be}{\begin{equation}}
\newcommand{\ee}{\end{equation}}
\newcommand{\bea}{\begin{eqnarray}}
\newcommand{\eea}{\end{eqnarray}}

\begin{document}
\draft

\title{On the relevance of center vortices to QCD}
\author{Philippe de Forcrand$^1$ and Massimo D'Elia$^2$}
\address{$^1$ETH-Z\"urich, CH-8092 Z\"urich, Switzerland}
\address{$^2$Dipartimento di Fisica dell'Universit\`a and I.N.F.N., I-56127, Pisa, Italy}

\date{\today}
\maketitle
 
\begin{abstract}
In a numerical experiment, we remove center vortices from an ensemble of 
lattice $SU(2)$ gauge configurations. This removal adds short-range disorder.
Nevertheless, we observe long-range order in the modified ensemble:
confinement is lost and chiral symmetry is restored (together with trivial
topology), proving that center vortices are responsible for {\em both} 
phenomena. As for the Abelian monopoles, they survive but their percolation 
properties are lost. 
\end{abstract}
 

The essential non-perturbative properties of QCD are confinement and chiral
symmetry breaking ($\chi$SB). It has been observed through numerical lattice simulations
that these two properties persist in the quenched theory 
up to a critical temperature 
$T_c \sim 220 MeV$, where confinement is lost; chiral symmetry appears to be 
simultaneously restored \cite{KS}. The disorder which leads to the area law for
the Wilson loop thus seems to be somehow tied to the existence of a chiral 
condensate. Although effective mechanisms have been proposed to
explain confinement or $\chi$SB,
no successful common explanation is yet available.

Two effective descriptions of QCD have been receiving a lot of attention:
one considers instantons as the effective degrees of freedom (d.o.f.), the other 
chromo-magnetic monopoles. 
Instantons are natural candidates to explain $\chi$SB: 
each instanton is associated with a zero mode of the Dirac operator \cite{tHooft},
and there must be an accumulation of zero eigenvalues to obtain a 
quark condensate \cite{BanksCasher}. Above $T_c$, the vanishing
of the condensate must 
\linebreak
correspond to a qualitative change in the instanton
ensemble (see, e.g., \cite{Shuryak}). On the other hand, it is unlikely that 
instantons play a significant role in confinement (see \cite{Diakonov} for a recent
discussion).
An attractive mechanism for confinement is
dual superconductivity of the QCD vacuum \cite{tHooftMandelstam}.
Considerable evidence for this dual Meissner effect has been accumulated
on the lattice, including a disorder parameter demonstrating 
the condensation of chromo-magnetic monopoles below $T_c$ \cite{DiGiacomo}.
This condensation has been observed directly
after gauge-fixing to ``Maximal Abelian Gauge'' \cite{Suzuki}, and
the idea of ``Abelian dominance'' has emerged, according to which the Abelian
d.o.f. of the Yang-Mills field encode all its long-distance (IR) properties.
Indeed, $\chi$SB and its restoration have been observed in the
Abelian sector \cite{Miyamura}.
The Abelian dominance scenario has some flaws, however:
it does not explain the breaking of the adjoint string,
and the Abelian string tension differs slightly from the Yang-Mills
\linebreak
one \cite{Bornyakov}. 
Moreover, d.o.f.
more elementary than Abelian monopoles, embedded in them and solely 
responsible for the physics assigned to them,
cannot be ruled out.

The idea of center vortices, which initially failed due to their misidentification 
\cite{Mack},
has been successfully revived, first as an embedded model inside the Abelian
sector \cite{Greensite}, then without
reference to Abelian projection \cite{Greensite_direct}. 
Center vortices are exposed by gauge-fixing:
after a gauge transformation which brings each lattice link as close as possible
to a center element of the gauge group, vortices consist of defects in the 
center-projected gauge
field. The idea of center dominance is
again that the center d.o.f. encode all the IR physics.
The density of center vortices seems to be a well-defined 
continuum quantity \cite{Greensite_direct,scaling}; 
the center string tension appears to more or less match the original one; 
and an explanation for the behaviour of the adjoint
potential has been proposed \cite{adjoint}. However, chiral symmetry has not yet been
studied in this context. 

An additional problem with Abelian and center dominance is that the relevant 
d.o.f. are identified only after gauge-fixing. Gauge-fixing non-Abelian fields
is notoriously ambiguous, and different Gribov copies produce
different Abelian monopoles or center vortices, whose properties like the string
tension differ slightly. 
For this reason, we are a priori suspicious of effective models which involve 
gauge-fixing, and so we designed a simple numerical experiment to disprove
the center-dominance scenario. For simplicity, we consider the gauge group $SU(2)$,
with center $Z_2$.
Our starting point is an ensemble of lattice gauge fields representative of the
continuum.
We identify center vortices in this ensemble, and construct from it a {\em modified} 
ensemble
where all center vortices have been removed by flipping
the sign of a subset of $SU(2)$ gauge links. This operation introduces
a lot of disorder in the gauge field.
Nonetheless, these disordered gauge fields now have a trivial, vortex-free 
center projection, and so according to the credo of center dominance
they should not confine. That is, our introduction of short-range disorder should
at the same time bring long-range order.
To our surprise, 
this is indeed what happens.
One may then ask if the 
spectral properties of the Dirac
operator are not also dominated by the center components of the gauge field. 
In that case, our modified ensemble should show no sign of $\chi$SB,
since its center projection is the trivial (perturbative) vacuum.
Indeed, this is what we observe: removal of center vortices causes {\em both}
loss of confinement and restoration of chiral symmetry.

The next intriguing question regards the fate of Abelian monopoles
as center vortices are removed. 
They do not disappear; on the 
contrary, the introduction of short-range disorder increases their number.
However, we observe the complete disappearance of monopole current
loops winding around the periodic lattice: we can thus identify
these as the fundamental objects associated with confinement
in the Abelian sector, apparently influenced by the underlying
center d.o.f.

Finally, we investigated the effect of Gribov copies which caused our initial
skepticism. We repeated our experiment on the same $SU(2)$ ensemble, but
introduced a systematic tolerance in the gauge condition to be satisfied
before identifying the center vortices to be removed.
Although the location and number of center vortices we removed varied appreciably,
the modified $SU(2)$ ensemble was always non-confining and chirally symmetric.

{\it The numerical experiment} -- We start from an ensemble of $SU(2)$ lattice gauge fields obtained by Monte Carlo
using the standard Wilson plaquette action. 
To identify center vortices, we gauge-fix our configurations in order to
bring each $SU(2)$ gauge link $U_\mu(x)$ as close as possible to an element of
the center $Z_2 = \{+{\bf 1},-{\bf 1}\}$. We therefore try to iteratively maximize
\be
Q(\{U_\mu\}) \equiv \sum_{x,\mu} \left( \mbox{Tr}~U_\mu(x) \right)^2 \; ,
\label{max}
\ee
as in \cite{Greensite_direct}, where this gauge is called the ``direct maximal center 
gauge.'' The gauge-fixed $SU(2)$ links, denoted $U_\mu^{GF}(x)$, are then
projected to $Z_2$ elements $Z_\mu(x)$ using
\be
Z_\mu(x) = \mbox{sign}[\mbox{Tr}~U_\mu^{GF}(x)] \; .
\label{cproj}
\ee
Plaquettes in the $Z_2$-projected theory with value $-1$ represent defects of
the $Z_2$ gauge field called P-vortices \cite{Greensite}.
Numerical evidence has been presented \cite{Greensite,Greensite_direct}
showing that plaquette-like P-vortices signal the presence of macroscopic, physical
excitations, called center vortices, in the unprojected original $SU(2)$ 
configuration.

Consider then the modified $SU(2)$ configuration made of gauge links $U'_\mu(x)$
constructed as
\be
U'_\mu(x) \equiv Z_\mu(x)~U_\mu(x)\; .
\label{flipdef}
\ee
The gauge transformation which maximizes $Q(\{U_\mu\})$ in (\ref{max}) also gives 
the same maximum to $Q(\{U'_\mu\})$, so that the modified gauge-fixed links
$U'^{GF}_\mu(x)$ are simply $Z_\mu(x)~U_\mu^{GF}(x)$. Therefore, we instantly
know the center projection $Z'_\mu(x)$ of $U'_\mu(x)$:
\bea
Z'_\mu(x) & = & \mbox{sign}[\mbox{Tr}~U'^{GF}_\mu(x)] = \mbox{sign}[\mbox{Tr}~Z_\mu(x)~U_\mu^{GF}(x)] \nonumber \\
& = & \left( \mbox{sign}[\mbox{Tr}~U_\mu^{GF}(x)] \right)^2 = {\bf +1} \; .
\label{flipped}
\eea
Every modified configuration $U'$ thus projects onto the trivial $Z_2$
vacuum: all center vortices have been removed.

Our ensemble consists of about 1000 $SU(2)$ configurations, on a $16^4$ lattice at
$\beta=2.4$. 
To maximize $Q(\{U_\mu\})$ (Eq.\ref{max}) we use standard overrelaxation,
stopping when
\be
\epsilon \equiv \sum_{x,\mu} \Delta (\mbox{Tr}~U_\mu(x))^2 < 10^{-6}
\label{stop_criterion}
\ee
from one gauge-fixing sweep to the next.

\begin{figure}[!ht]
\begin{center}
\vspace{-2.0cm}
\epsfig{file=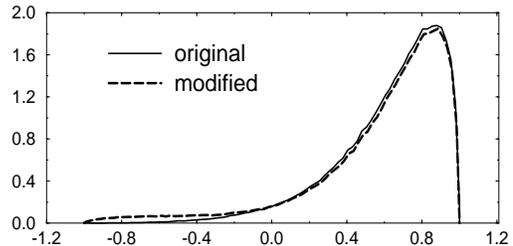,height=7.0cm,width=6.5cm,angle=-90}
\vspace{-1.0cm}
\caption{Normalized plaquette distribution on the original and
modified ensembles ($SU(2)$, $\beta = 2.40$).
Center-vortex removal increases short-range disorder.}
\end{center}
\end{figure} 
\vspace{-0.5cm}

In Fig.1 we show the distribution of $SU(2)$ plaquette values on the original
and modified ensembles. It is apparent that under the sign flip Eq.(\ref{flipdef}),
many $SU(2)$ plaquettes acquire a negative value. 
The modified ensemble has an increased action, i.e. more short-range disorder.

\begin{figure}[!ht]
\begin{center}
\vspace{-1.0cm}
\epsfig{file=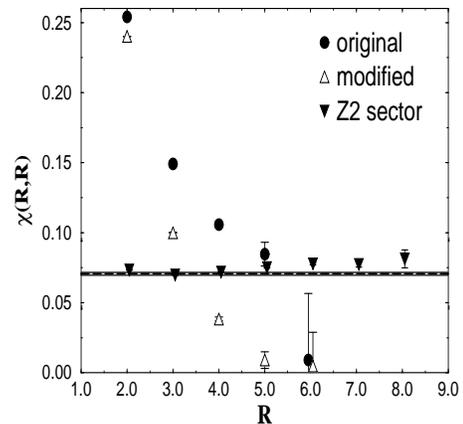,height=6.5cm,width=7.5cm,angle=-90}
\caption{Creutz ratios 
on the original and modified ensembles. The dashed band is the string
tension result of \protect\cite{Teper}.}
\end{center}
\end{figure} 
\vspace{-0.5cm}

{\it Results} -- In Fig.2 we present our results for the Creutz ratios 
$\chi_{R,R} \equiv - \ln [\langle W_{R,R} \rangle 
\langle W_{R-1,R-1} \rangle
/\langle W_{R,R-1} \rangle^2]$
constructed from averages $\langle W_{R,T} \rangle$ of $R$ by $T$ Wilson loops
on the original and modified ensembles. For large $R$, $\chi_{R,R}$ tends to the
string tension $\sigma$. 
On the modified ensemble, the Creutz
ratios clearly decrease and tend to zero. 
Despite the increased short-range disorder,
long-range order has been created and confinement has been lost.

\begin{figure}[!th]
\begin{center}
\epsfig{file=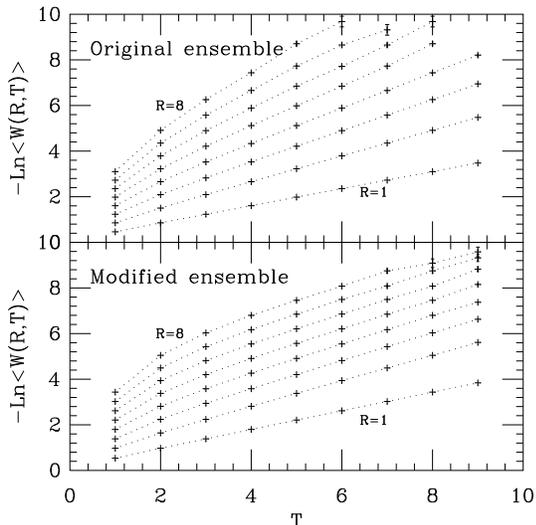,height=7cm,width=7cm}
\caption{Wilson loop values $\langle W_{R,T} \rangle$
on the original and modified ensembles.
Note the parallel lines for successive $R$ in the latter:
upon center-vortex removal, confinement is lost.}
\end{center}
\end{figure} 
\vspace{-0.5cm}

This is even clearer if one looks directly at the Wilson loop values. In Fig.3
we show $- \ln \langle W_{R,T} \rangle$ as a function of $T$.
For a fixed $R$, points at successively larger $T$ form a line whose 
asymptotic slope is the value of the static potential $V(R)$. 
The lines corresponding to the modified ensemble are parallel, indicating
that $V(R)$ does not grow with $R$: the string tension has vanished.

\begin{figure}[!h]
\begin{center}
\epsfig{file=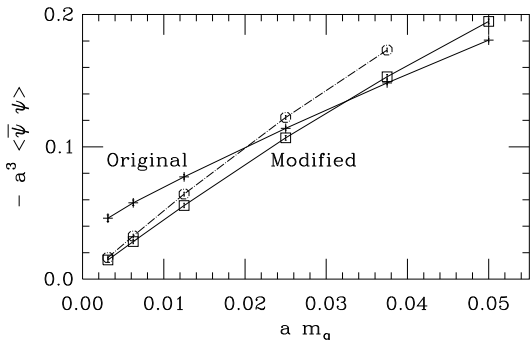,height=4.5cm,width=7cm}
\caption{Quark condensate $\langle \bar{\psi} \psi \rangle(m_q)$ on the
original and modified ensembles. The dashed line corresponds to ``poor''
center-vortex identification (see text). In all cases,
center-vortex removal restores chiral symmetry.}
\end{center}
\end{figure} 
\vspace{-0.5cm}

Fig.4 illustrates our study of chiral symmetry on the original and modified
ensembles. As is well-known, $\chi$SB cannot occur on a 
finite lattice. Therefore, we measure 
$\langle \bar{\psi} \psi \rangle(m_q) = 
\langle \mbox{Tr} (/\!\!\!\!D + m_q)^{-1} \rangle$
for a range of quark masses $m_q$ where finite-size effects are small,
and extrapolate to $m_q \rightarrow 0$.
In the original ensemble, $\langle \bar{\psi} \psi \rangle$ clearly
extrapolates to a non-zero value which signals $\chi$SB.
In the modified ensemble, the extrapolated value is zero within errors:
center-vortex removal restores chiral symmetry.
We expect then the instanton content of the Yang-Mills field to
be modified also. To check this, we use improved cooling \cite{imp_cool}
to measure the topological charge of the modified field:
the striking result is that the removal of center
vortices always leads to the trivial topological sector.

We therefore have clear evidence for ``center dominance'':
in our modified ensemble, where the center-projected field is the trivial
vacuum (all links equal to ${\bf 1}$), 
the Yang-Mills field shows
the IR properties of the trivial vacuum, i.e., no confinement,
no $\chi$SB and no topology. The IR properties of the Yang-Mills
field appear to be determined by its center projection.

On the other hand, a large number of studies now support the alternative
scenario of ``Abelian dominance.'' We use our approach of center-vortex
removal to directly assess the relationship between these two scenarios.

In a first experiment, we construct the Abelian projection of our original
$SU(2)$ ensemble by gauge-fixing 
to Maximal Abelian Gauge in the usual way
\cite{Suzuki}, then identify and remove center vortices from the {\em Abelian}
sector. While the original Abelian-projected ensemble shows confinement, with
a string tension similar to the non-Abelian one, the modified Abelian-projected 
configurations do not confine.
Therefore, we find no contradiction between ``Abelian dominance'' and
``center dominance.'' The latter simply appears more fundamental because
of the greater reduction of the number of d.o.f.

\begin{figure}[!th]
\begin{center}
\vspace{-2.0cm}
\epsfig{file=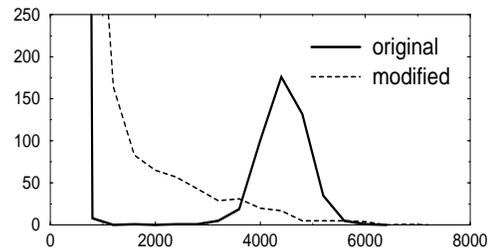,height=6.5cm,width=6.5cm,angle=-90}
\vspace{-1.0cm}
\caption{Size distribution of monopole clusters 
on the original and modified ensembles.}
\end{center}
\end{figure} 
\vspace{-0.5cm}

In a second experiment, we look at clusters of Abelian monopole currents,
whose percolation has been identified as the signal for confinement
\cite{Polikarpov},
obtained from the original and modified ensembles.
We find that the removal of center vortices changes the distribution of monopole
cluster sizes in a crucial way (see Fig.5) : whereas in the original ensemble, each 
configuration contains typically one very large, percolating monopole cluster
and many very small ones, the modified ensemble gives a more homogeneous
size distribution, with a handful of large clusters per configuration;
these are the remnants of the very large one, broken into pieces
by the vortex removal.
Some of them still percolate, even though confinement has disappeared.
Therefore, we are led to associate confinement with a more specific feature of
the monopole clusters: monopole current loops which wind around the periodic lattice.
Such loops can be found frequently on the original, confining ensemble,
but never on the modified, non-confining one. 
We conclude that: $(i)$ on a finite lattice confinement manifests itself in the Abelian sector by the
presence of monopole current loops with non-trivial topology;
$(ii)$ center-vortex removal, which destroys confinement, always finds the
``weak links'' of these non-trivial loops and breaks them into trivial pieces.

Now let us consider the issue of gauge-fixing ambiguities, which was the reason for
our initial skepticism about the center-vortex idea. These ambiguities come from the
structure of $Q$ (Eq.\ref{max}), which has many local maxima, any of which can
be selected by a local iterative maximization algorithm. Each local maximum,
or Gribov copy, will have its own set of P-vortices, differing in number and location.
The proposal of \cite{Greensite} is that, no matter which Gribov copy one chooses,
P-vortices are the traces of physical center vortices and are roughly located
at their center. This argument may account for P-vortices differing in location
but not in number. To study this question in more
detail, we magnified the effect of gauge-fixing ambiguities, by stopping our
iterative algorithm early, as soon as $\epsilon$ (Eq.\ref{stop_criterion}) $< 10^{2}$.
Thus we not only explore a different basin of attraction of $Q$, but we do not even
stop at a local maximum. 
One effect of this partial gauge-fixing is expected:
the density of P-vortices increases from $\rho \approx 5.5\%$ to $\approx 7.4\%$, 
i.e. shows an increase $\delta\rho \approx 1.9\%$.
The string 
\linebreak
tension measured in the $Z_2$-projected ensemble
increases accordingly: whereas the $Z_2$ string tension after ``complete'' gauge-fixing 
($\sigma a^2 \sim 0.075$) is a little larger than 
\linebreak
but compatible with
the non-Abelian string tension ($0.0708(11)$ \cite{Teper}, see Fig.2),
it jumps to $\approx 0.12$ after partial gauge-fixing.
What is remarkable is that this increase $\delta\sigma \approx 0.045$ is 
similar to that obtained by placing the surplus $\delta\rho$ of P-vortices 
at {\em random}, uncorrelated locations: 
$\delta\sigma \approx -\ln(1 - 2 \delta\rho)$.
This makes it plausible that the center projection always
captures the core d.o.f. relevant for IR properties,
plus a varying amount of unrelated noise \cite{footnote}.
Indeed, it has been argued that
gauge-fixing is not even necessary for center projection \cite{Greensite_last}.
In our description, center-gauge-fixing acts as a UV noise-filtering device,
with different Gribov copies letting through different noise components.
Further evidence for this is obtained by removing from the original $SU(2)$ 
ensemble the center vortices identified after partial gauge-fixing only.
Just as for ``complete'' gauge-fixing, we observe that confinement is lost,
chiral symmetry restored, and the topology trivial.
The only difference is that the modified ensemble now has much more short-range
disorder.

{\em In conclusion}, we have shown that removal of center-vortices from an $SU(2)$
Yang-Mills ensemble causes the loss of confinement {\em and} the restoration of chiral
symmetry. 
One may ask about the connection of the modified ensemble $\{U'\}$
to the physics of the original $SU(2)$ theory. Note that only plaquettes 
of  $\{U'\}$ at the locations of P-vortices differ from those
of $\{U\}$: hence, as $a \to 0$, their proportion goes to zero as $a^2$,
since the density of P-vortices is physical \cite{Greensite_direct,scaling}.
Therefore, rewriting Eq.\ref{flipdef} as 
$U_\mu(x) \equiv Z_\mu(x)~\times~U'_\mu(x)$, we see that the original 
field $\{U\}$ has been factorized into a (maximally)
central part $\{Z\}$ and a quotient, $\{U'\}$, whose field strength
differs from the original one only on defects of codimension 2.
Nevertheless, this 
small difference alters the physics dramatically: $\{U'\}$ has perturbative
properties, so that all the non-perturbative, IR physics
{\em must} be carried by $\{Z\}$, which by definition encodes the center
vortices.
It would be desirable, of course, to formulate an effective action for the
center-projected theory. Ref.\cite{Engelhardt} considers an extension
of the Nambu-Goto action, where the fundamental d.o.f. are the 2-dimensional random
surfaces dual to the 
\linebreak
P-vortices. Ref.\cite{Polikarpov_Z2} instead proposes to
consider center monopoles and their world-lines.
We suggest identifying a ``minimum spanning tree'' of negative $Z_2$ links responsible for
the P-vortices: perhaps only a subset of them form the essential d.o.f. 
governing the IR properties.

Finally, our vortex-removal procedure can be used to study properties of
non-confining non-Abelian fields and effects of center-symmetry breaking. 
For instance, removing time-like center disorder
only would be similar to raising the temperature above $T_c$.

\par\bigskip
    
We thank A. Di Giacomo, M. Golterman, J. Greensite, T. Kovacs, C. Lang and 
O. Miyamura for discussions.


\begin{thebibliography}{99}
 
\bibitem{KS} J. Kogut et al., Phys. Rev. Lett. 50 (1983) 393.
\bibitem{tHooft} G. t'Hooft, Phys. Rev. D14 (1976) 3432.
\bibitem{BanksCasher} T. Banks, A. Casher, Nucl. Phys. B169 (1980) 103.
\bibitem{Shuryak} T. Sch\"afer, E.V. Shuryak, hep-ph/9610451.
\bibitem{Diakonov} R.C. Brower et al., hep-lat/9809091.
\bibitem{tHooftMandelstam} G. t'Hooft, Nucl. Phys. B190 (1981) 455; 
S. Mandelstam, Phys. Rep. 23C (1976) 245.
\bibitem{DiGiacomo} A. Di Giacomo, G. Paffuti, Phys. Rev. D56 (1997) 6816.
\bibitem{Suzuki} H. Shiba, T. Suzuki, Phys. Lett. B333 (1994) 461.
\bibitem{Miyamura} O. Miyamura, Phys. Lett. B353 (1995) 91.
\bibitem{Bornyakov} G.S. Bali et al., Phys. Rev. D54 (1996) 2863.      
\bibitem{Mack} G. Mack, E. Pietarinen, Nucl. Phys. B205 (1982) 141.
\bibitem{Greensite} L. Del Debbio et al., Phys. Rev. D55 (1997) 2298.  
\bibitem{Greensite_direct} L. Del Debbio et al., Phys. Rev. D58 (1998) 094501.
\bibitem{scaling} K. Langfeld et al., Phys. Lett. B419 (1998) 317.
\bibitem{adjoint} M. Faber et al., Phys. Rev. D57 (1998) 2603.
\bibitem{Teper} C. Michael, M. Teper, Phys. Lett. B199 (1987) 95.
\bibitem{imp_cool} P. de Forcrand et al., Nucl. Phys. B499 (1997) 409.
\bibitem{Polikarpov} E.g. T.L. Ivanenko et al., Phys. Lett. B302 (1993) 458.
\bibitem{Greensite_last} M. Faber et al., hep-lat/9810008.
\bibitem{footnote}
This is what seems to happen
for a generic starting point on the gauge orbit. There also exist
pathological starting points, notably Landau gauge, from which one obtains very 
few center vortices, not associated with confinement. 
\bibitem{Engelhardt} M. Engelhardt, H. Reinhardt, UNITU-THEP-3-99. 
\bibitem{Polikarpov_Z2} M.N. Chernodub et al., hep-lat/9809158.

 
\end{thebibliography}
\end{document}